# Weakest Preconditions and Cumulative Subgoal Fulfillment: A Comparison

**Abstract**

We contrast the use of weakest preconditions for the correct construction of procedures with the cumulative subgoal fulfillment (CSF) approach. An example of Cohen and Monin is used for this purpose. The CSF construction process is demonstrated.

## 1. Introduction

A motive of Dijkstra's approach [Di] and that of many others (e.g., Floyd[Fl] and Gries [Gr]) has been to make correctness integral with program construction. This includes the use of *weakest preconditions*. We compare the latter with *cumulative subgoal fulfillment* (CSF), introduced by the author in [Br] and used in [BK] and [KPB].

In [Co], Cohen showed how to use weakest preconditions to compute $N^3$ without the use of multiplication. His solution, a very short O(*n*) program shown below, is described by Monin [Mo] as "striking." Monin simplified its description. Cohen's demonstration is indeed striking but its creation is a lengthy, somewhat ad hoc process. In this paper, we describe the CSF approach (Sections 2 and 4), place it in the context of related research (Section 4), and apply it to the $N^3$ problem (Sections 5, 6, and 7). CSF is a systematic process that yields several solutions, including Cohen/Monin's, as well as O(log *n*) solutions (Section 8).

## 2. CSF

CSF is an approach to creating procedures hand-in-hand with correctness. It is based on ideas of Dijkstra, Hoare, et al. but introduces the notion of *accumulation*, a principle borrowed from physical construction. In the latter, the completion of each part can be thought of as the fulfillment of a subgoal (or intermediate goal)–which typically remains valid while additional parts are built. For example, once a suspension bridge tower has been constructed, it is expected to remain intact. Correspondingly, CSF consists of a sequence of code blocks, each of which fulfills a subgoal, and which leaves invariant subgoals already fulfilled. CSF has been, and is being applied to a broad array of problems.

## Definition of CSF

The definition of CSF is as follows. Let *P* be a procedure for which *pre*, *inv*, and *post* are the conjunctions of its preconditions, invariants, and postconditions respectively. An *algorithm plan* for *P* is a sufficient sequence $s_1, s_2, ..., s_n$ of predicates i.e., one satisfying the following.

$pre \wedge s_1 \wedge s_2 \wedge ... \wedge s_n \wedge inv \Rightarrow post \wedge inv$.

An example is *int getMax( int[] anArr )*, as follows.

    *pre*:    *anArr.length* >= 1

    *post*:   *returnI = anArr[r]* for some 0 <= r < *anArr.length*
           AND *returnI* >= *anArr[i]* for *i* = 0, ..., *anArr.length* - 1

    $s_1$:    0 <= r < *anArr.length*  AND *returnI = anArr[r]*
           AND *i* < *anArr.length*  AND *returnI* >= *anArr[j]* for *j* = 0, ..., *i*

    $s_2$:    *i* = *anArr.length* - 1

A CSF implementation of *P* consists of an algorithm plan $s_1, s_2, ..., s_n$ and a sequence $c_1, c_2, ..., c_n$ of code blocks satisfying the following Hoare triples.

    (1) (Preconditions applicable): $inv \wedge pre$ {$c_1$} $inv \wedge s_1$, and

    (2) (Subgoals accumulate):    for *i* =2, 3, ..., *n*, $inv \wedge s_1 \wedge s_2 \wedge ... \wedge s_{i-1}$ { $c_i$ } $inv \wedge s_1 \wedge s_2 \wedge ... \wedge s_i$

For the *getMax*() example, we can take $c_1$ as "i, returnI := 0, anArr[0];" and $c_2$ as "**while**(i < anArr.length) **do** ... **od**," where each iteration of the loop restores $s_1$.

## Terminology, Remarks, and Notation

A predicate *p* which is consistent with the postconditions (i.e., for which $p \wedge post =$ **true**) will be called a *cumulative subgoal* for *P*. An algorithm plan therefore consists of cumulative subgoals.

A subgoal in an algorithm plan whose presence is logically redundant (though presumably useful) will be called *pragmatic* i.e., a member *s* of an algorithm plan *P* for which *P\s* is also an algorithm plan for the same invariants, pre- and postconditions. For convenience in verifying that a set of subgoals is sufficient, we sometimes prepend a subgoal's label with a bracket to note it as pragmatic, as in [SG3. For convenience in maintaining invariance, we sometimes append to a subgoal's label an angle bracket when its fulfillment involves constants only. For example, although the following subgoal is an essential part of an algorithm plan below, it can be ignored when checking the sufficiency of subgoals, and it need not be revisited for restoration.

[SG> (Square): $s = N^2$

When it is possible to easily fulfill subgoals $s_i$ and $s_j$ via a single block of code, we generally do so, even though this does not, strictly speaking, follow the definition of CSF. This fits with no difficulties within a CSF implementation. However, $s_i$ and $s_j$ are generally restored separately. (Indeed, if $s_i$ and $s_j$ were initially fulfilled, and always restored, via single blocks of code, we would substitute them with the single subgoal $s_i \wedge s_j$.)

Sequence, branch, and loop have long been recognized as fundamental constructs of programming (Böhm and Jacopini [BJ]). The use of CSF may appear to produce only sequences but this is not the case. The number of subgoals may be determined at runtime ("Version 1" below is an example), and so may indeed form an outer loop. Secondly, branching, as in "**if** *Condition* **then** *A* **else** *B*" fulfills the following pair of cumulative subgoals.

Subgoal 1. (*Condition* $\wedge$ *A*) $\vee$ $\rceil$*Condition*
Subgoal 2. ($\rceil$*Condition* $\wedge$ *B*) $\vee$ *Condition*

In CSF, subgoals are declarative whereas loops, branches etc. are program constructs employed to fulfill them.

In this paper, *'x* (or $\underline{x}$) and *x'* refer to the value of *x* before and after the relevant operation, respectively.

## 3. CSF Patterns

This section describes common CSF patterns used in this paper.

### *Standard* and *Possession* Subgoal Creation

There are many ways in which subgoals can be created. The most common, used several times in this paper, is a technique for selecting loop invariants made clear by Gries [Gr], viz. to *replace a constant with a variable*. This will be referred to as the "standard method" for subgoal creation. An example is to replace the size of an array with an index.

The second subgoal creation technique used in this paper is to possess a useful expression. Naming such an expression is an old idea in programming (and an ancient corresponding one in mathematics,) but when it is used as a CSF subgoal, there is an obligation to maintain the relationship. We will refer to this as a *possession* subgoal.

### *Standard Fulfillment* Pattern

In most cases, the code blocks $c_1, c_2, c_3, \ldots$ referred to in Section 2 can be constructed via the following *fulfill/restore* sequence. For restorations performed serially, the order of initial subgoal fulfillment suggests the same order for restoration.

```
// Fulfill s₁: (i.e., the following is c₁)
    <use pre to fulfill s₁>
    <restore inv>
// Fulfill s₂
    <use inv and s₁ to fulfill s₂>
    <restore inv and s₁ (in sequence or in parallel)>
// Fulfill s₃
    <use inv, s₁, and s₂ to fulfill s₃>
    <restore inv, s₁ and s₂>...
```

This "standard fulfillment" pattern facilitates code comprehension.

### *Standard Incremental Fulfillment* Pattern

A variation on the standard fulfillment pattern occurs when subgoal $s_i$ is fulfilled incrementally. In that case, $c_i$ consists typically of a repeated perturb/restore loop as follows.

```
while( !sᵢ ) do  // sᵢ on termination
    <perturb productively>
    <using inv, s₁, s₂, ... , and sᵢ₋₁ (i.e., values prior to perturbation),
      restore inv and s₁-sᵢ₋₁ in sequence or in parallel>
od  // <proof of termination ...>
```

A *productive* perturbation is one that eventually moves the loop toward termination.

Applying this to the *getMax*() example above, $c_2$ would be as follows.

```
while( i < anArr.length ) do  // s2 on termination
    ++i; // productive perturbation
    // Restore s1
    if( anArr[i] > returnI )
        returnI := anArr[i];
od  // terminates because ...
```

In the standard incremental fulfillment pattern, $inv \wedge s_1 \wedge s_2 \wedge \ldots \wedge s_{i-1}$ is a loop invariant. CSF does not actually require such a loop invariant – only that the last iteration restores $inv \wedge s_1 \wedge s_2 \wedge \ldots \wedge s_{i-1}$.

## 4. Relationship of CSF with Existing Work

The purpose of this paper is to compare CSF with the use of weakest preconditions. This section places CSF in context with other related research.

*Invariance* is a significant part of CSF, being key to accumulation. An important early reference to invariance is Floyd [Fl]. For a time, invariance was thought of largely in the context of loops but this has given way to a wider appreciation of its applicability. For example, in a series of papers, Ernst et al ([Er1], [Er2], [Er3]) demonstrated the detection of invariants at places in as-built code.

CSF relies on *Design by Contract* (DbC), which has a rich literature and application base (see, for example [Me]). In a sense, CSF extends DbC by instituting "contracts" for subgoals—not just for procedure goals.

The objectives of CSF are close to those of Naur's *Action Clusters* [Na1] whose description he included in his 2006 Turing Lecture [Na2]. However, CSF differs from Action Clusters in that it is simpler and standardized. Action Clusters do not explicitly accumulate fulfilled subgoals.

The cumulative nature of CSF is consistent with the monotonic philosophy of agile programming. As pointed out by Martin [Ma], agile processes try to observe Liskov's *Open-Closed Principle* in which code is amenable to extension but not necessarily to modification.

Cumulative Subgoal Fulfillment is complementary to *Hoare Logic* proofs. CSF facilitates the creation of code by decomposing specifications into smaller, more manageable sub-specifications i.e., subgoals. Once a subgoal is cumulatively fulfilled in code, formal verification via Hoare logic can be applied to it. *Theorem provers* can be used to verify that a cumulative implementation of a subgoal is correct; they can also be used to verify that an algorithm plan is sufficient for postconditions.

*Refinement* is a series of conversions to increasingly specific forms, starting with postconditions, and ending with an implementation. All methodologies begin with specifications and end with implementation. The CSF process—identifying cumulative subgoals—is not particularly a sequence of increasingly specific forms.

## 5. Cohen/Monin's Solution

The following is Monin's simplification of Cohen's solution, where *c* holds the desired $N^3$.

```
r,d,c,e := 0,0,1,6;
while r≠N do
    r,c,d,e := r+1, c+d, d+e, e+6 od
```

This is a surprisingly compact program but its synthesis and correctness proof, based on weakest preconditions, require two pages of separate, specific explanation. Table 1 contains some of this explanation.

## 6. O($n$) CSF Solutions

To synthesize a solution via CSF, we create cumulative subgoals. We will show five ways to do this, depending on the programmer's motives. Three of these, of O($n$), are in this section and two, of O($n^2$), are in Section 8.

### Version 1: A Transparent O($n$) Solution

With $c=N^3$ as the postcondition, a simple approach is to express this as $c = N^2 \cdot N$. This requires the pragmatic possession subgoal $s=N^2$, yielding the following algorithm plan.

    [SG1>      $s = N^2$

    SG2      $c=N^3$

Each of these can be fulfilled via a loop with $N$ iterations, where the fulfillment of SG2 uses SG1. This approach remains viable for computing $N^M$ in O($MN$) time via an algorithm plan containing $M$-1 subgoals.

### Version 2: An O($n$) Solution Easily Derived

For this application of CSF, we begin with the overall goal, $c=N^3$, then use the standard subgoal creation technique, replacing the constant $N$ with a variable $r$. This is the same starting point as Cohen/Monin. The following algorithm plan results.

    SG (Cube)      $c = r^3$ AND $r \leq N$

    SG (Particularized)      $r = N$

The assignments "$c, r := 0, 0;$" fulfills subgoal "Cube." The standard incremental fulfillment pattern described in Section 2 fulfills "Particularized" as follows.

```
while r ≠ N do              // termination fulfills "Particularized"
      r := r+1;             // productive perturbation
      <Restore "Cube">
od                          // (terminates because ...)
```

Restoring the "Cube" subgoal means establishing $\underline{c} = \underline{r}^3$. The latter is $(\underline{r}+1)^3 = \underline{r}^3 + 3\underline{r}^2 + 3\underline{r} + 1$. The quantity $\underline{r}^3$ is just $\underline{c}$ and it is easy to obtain $3\underline{r} + 1$. We thus need the following pragmatic possession subgoal.

    SG (Square): $s = r^2$

This yields the CSF implementation below. The first two subgoals are easily fulfilled together (but restored separately). We use the standard fulfillment pattern overall, and the standard incremental pattern to fulfill SG2, described in Section 3. In the code, the addition of a fixed number *p* of quantity q can be abbreviated p*q without compromising the stricture on multiplication assumed by this paper e.g., "2*r" below is shorthand for "r+r."

    //----SG1 (Cube):                  $c = r\hat{\ }3$ AND $r \leq N$
         /* fulfilled together with */
    //---[SG2 (Square):              $s = r\hat{\ }2$

    r, c , s = 0, 0, 0;

    //----SG3 (Particularized):    $r = N$

    **while** r≠N **do**  // perturb productively; restore c and s
        r, c, s := r+1, c + 3*(s+r) + 1, s + 2*r + 1 **od**

## Version 3: The Cohen/Monin Solution

Instead of possessing merely $r^2$, as in Version 1, one can streamline the restoration of *c*. Since $c' = \underline{r}^3 + 3\underline{r}^2 + 3\underline{r} + 1$, we introduce the following possession subgoal.

    SG (Quadratic): $q = 3r^2 + 3r + 1$

This is simple to fulfill. Its restoration requires $3(\underline{r}+1)^2 + 3(\underline{r}+1) + 1 = 3\underline{r}^2 + 9\underline{r} + 7$, which is $\underline{q} + 6\underline{r} + 6$. Continuing the process, we add the following possession subgoal.

    SG (Linear): $l = 6r + 6$

The resulting CSF program is as follows. The first three subgoals can be fulfilled together (but restored separately).

    //----SG1 (Cube)                $c = r^3$ AND $r \leq N$
         /* fulfilled together with */
    //---[SG2 (Quadratic)        $q = 3r^2 + 3r + 1$
         /* fulfilled together with */
    //---[SG3 (Linear)             $l = 6r + 6$

```
            r, c, q, l = 0, 0, 1, 6;

    //----SG4 (Particularized)      r = N

          while r < N do                    // SG4 on termination
              r,c,q,l := r+1, c+q, q+l, l+6;  // perturbation; SG1-3 restored
```

This is the same implementation as Cohen/Monin.

## 7. Comparison of Weakest Precondition vs. CSF

Table 1 compares the weakest precondition process with CSF version 2. The quotes are from Monin [Mo]. Weakest precondition usage depends on programming constructs whereas CSF is driven by subgoals, which are usually derived from the postconditions.

| Step | Weakest Precondition [Mo] | CSF |
|------|---------------------------|-----|
|      | Postcondition: $c = N^3$ ||
| 1 | "Aiming at a loop" | Identify a sufficient set of subgoals |
| 2 | Replace a constant with a variable | Replace $N$ in $c = N^3$ with $r$, obtaining SG1 (Cube): $c = r^3$ AND $r \leq N$ |
| 3 | "Put the postcondition in the form $I \wedge \neg C$. The only available constant is $N$, hence we put the postcondition in the form $C = r^3 \wedge r = N$" | Complete an algorithm plan. SG2 (Particularized): $r = N$ |
| 4 | (Initialization) | Fulfill SG1 (Cube): $c, r = 0, 0$ |
| 5 | "Look for a program having the following shape: <br> 1. 'Establish $I$' <br> 2. **while** $r \neq N$ **do** <br> 3. 'Preserve I while making $r$ closer to $N$' <br> where the loop invariant is $I$: $c = r^3$" | Since $\underline{c} = \underline{r}^3 + 3\underline{r}^2 + 3\underline{r} + 1$, we introduce the following possession subgoal. SG (Quadratic): $q = 3r^2 + 3r + 1$ |
| 6 | "The loop body contains $r=r+1$ and an assignment to c such that the invariant is preserved." | This is simple to fulfill. Its restoration requires $3(\underline{r}+1)^2 + 3(\underline{r}+1) + 1 = 3\underline{r}^2 + 9\underline{r} + 7$, or $\underline{q} + 6\underline{r} + 6$. Continuing the process, we add the following possession subgoal. SG (Linear): $l = 6r + 6$ |
| 7 | "The shape we envisage for line 3 is $r,c = r+1, E$ Where $E$ is an expression that is yet to be found, and we want $I \wedge r \neq N \Rightarrow [r, c = r+1, E]I$" | |
| 8 | <Several weakest precondition transformations are described here (in about half a page), yielding the following. $E = c + 3r^2 + 3r + 1$> | |

| 9 | "This raises a problem: it is not a sum of known quantities. Let us introduce *d* and assume, at the same time, that $d = 3r^2 + 3r + 1$" | |
| --- | --- | --- |
| 10 | "We actually consider *I* defined as $I_1 \wedge I_2$ ..." <resolved via a page of transformations> .... | |
| 11 | | Program |

Table 1: A Comparison of Weakest Preconditions and CSF for N Cubed Problem

## 8. O(log *n*) CSF Solutions

To improve efficiency, we look for more productive perturbations for the loop. Instead of perturbing *r* by 1, we attempt to double it. More precisely, we work with its binary representation.

Note first that a function *getBin*(), with the following specification and algorithm plan, creates the binary representation of *N* in O(log *N*) time since each subgoal can be fulfilled in at most O(log *N*) time.

    Post   (*N* in binary)         $N = b[0] + 2b[1] + ... + 2^k \cdot b[k]$
                                                AND $b[k] = 1$ AND $b[i] = 0$ or 1 for $0 \le i < k$

    SG1> (Log *N*)                $2^k \le N \le 2^{k+1}$
    [SG2> (Powers of 2)        $t = \{0, 2, 4, ..., k\}$
    SG3  (Tail of binary)        $N = m + 2^j b[j] + 2^{j+1} b[j+1] + ... + 2^k \cdot b[k]$
    SG4  (Complete)           $j=0$ AND $m=0$

SG3 can be fulfilled with "j, b[k], j, m := k, 1, N-t[k];" and SG4 can be fulfilled with *k* subtractions of t[j] from m.

### Version 4: A Transparent O(log *n*) Solution

Note next that once *N*'s binary representation is known, *nN* can be computed via addition alone in O(log *N*) time. For example, $n \cdot (101) = n \cdot 1 + (n+n) \cdot 0 + ((n+n) + (n+n)) \cdot 1$. We will assume that the following function performs this multiplication.

    addArg1Arg2Times( int intToAdd, int[] numTimes )

The following is then an O(*N*) cube function.

```
//---[SG1 (N in binary): ...

        b := getBin(N);

    //---[SG2 (Square): s = N²

        s := addArg1Arg2Times(N, b);

    //----SG3 (Cube): c == N³

        c := addArg1Arg2Times(s, b);
```

This approach remains viable for computing each $N^M$ in $O(M \log N)$ time.

## Version 5: A Cohen/Monin-Like O(log n) Solution

Although Version 4 is efficient, it lacks an interesting form like that of Version 3. To that end we calculate $(r + 2^j b_j)^3$. In the following, $j$ has already been incremented and $b_z$ is used interchangeably with $b[z]$. Note that $b_j^n = b_j$.

$$(\underline{r} + 2^j b_j)^3 = \underline{c} + b_j(3\underline{r}^2 \cdot 2^j + 3\underline{r} \cdot 2^{2j} + 2^{3j}) = \underline{c} + b_j(6 \cdot 2^{j-1}\underline{r}^2 + 12 \cdot 2^{2(j-1)}\underline{r} + 2^{3j})$$

We thus introduce the following possession subgoals.

SG (Square)   $s = 2^j r^2$

SG (Linear)   $l = 2^{2j} r$

To restore the "Square" subgoal, we calculate the following.

$$2^j(\underline{r} + 2^j b_j)^2 = 2^j \underline{r}^2 + 2^{2j+1} b_j \underline{r} + 2^{3j} b_j = 2 \cdot 2^{j-1} \underline{r}^2 + 8 \cdot 2^{2(j-1)} b_j \underline{r} + 2^{3j} b_j = 2s + b_j(8\underline{l} + 2^{3j})$$

To restore "Linear", we note the following.

$$2^{2j}(\underline{r} + 2^j b_j) = 2^{2j}\underline{r} + 2^{3j} b_j = 4 \cdot 2^{2(j-1)}\underline{r} + 2^{3j} b_j = 4\underline{l} + 2^{3j} b_j$$

The following algorithm plan results.

SG1> (N in binary)   $N = b[0] + 2b[1] + ... + 2^k b[k]$
                     AND $k[i] = 0$ or $1$ for $0 <= i < k$ AND $b[k] = 1$

[SG2> (Powers of 2)  $t = \{1, 2, 4, ..., 2^{3k+1}\}$

SG3 (Cube)           $c = r^3$ where $r = b[0] + 2b[1] + ... + 2^j b[j]$ and $j <= k$

[SG4 (Square term)   $s = 2^j r^2$

[SG5 (Linear term)    $l = 2^{2j}r$

SG6 (Particularized)    $j = k$

SG3-5 can be fulfilled with the following.

j, c, s, $l$ := 0, b[0] , b[0], b[0];

SG6 can be fulfilled as follows.

```
while j<k do                // SG6 on termination
    j := j+1;               // productive perturbation
    u, v = b[j], t[3*j];    // convenience
    // Restore SG3-5 (multiplication by u is allowed)
    c, s, l := c + u*(6*s + 12*l + v ), 2*s + u*(8*l + v), 4*l + u*v;
od
```

## 9. Summary and Conclusion

Monin [Mo] described and clarified an interesting example of Cohen's [Co] that applies weakest preconditions to compute the cube of a number in O($n$) time using only addition. A separate, lengthy explanation is required of its correctness and derivation. This paper showed that the Cumulative Subgoal Fulfillment (CSF) technique produces the same code in short order and in a standard fashion. CSF can also be used to produce O(log $n$) solutions, some of which are generalizable.